# Jupiter's North Equatorial Belt and Jet:
# II.  Acceleration of the jet and the NEB Fade in 2011-12

John H. Rogers                                             *A Report of the Jupiter Section*
___________________________________________________________________________

**Summary**

Paper I described the normal features of the North Equatorial Belt (NEB) in recent years, especially the large dark formations which are thought to represent waves in the prograde jet on the NEB south edge (NEBs, 7ºN), and the NEB expansion events (NEEs) in which the belt broadens to the north at intervals of 3 to 5 years.  Here I describe an exceptional set of apparently coordinated changes which occurred in 2011-12, after more localised precursors in 2008 and 2010.
(1) The large NEBs dark formations progressively disappeared until none remained.
(2) In the sectors of NEBs thus vacated, smaller dark features all moved with unprecedented 'super-fast' speeds, which were modulated by the few normal features as long as they lasted, and then accelerated further, reaching 139-151 m/s.
(3) 'Rifts' (expanding systems of convective white clouds) also disappeared from the NEB.
(4) The NEB north half progressively faded (whitened) until there was only a narrow, southerly NEB, narrower and fainter than it had been for nearly a century.
These changes have several profound implications for understanding the dynamics of the region.  First, the NEBs took on the same appearance, dynamics, and speed, as the equivalent jet at 7ºS (SEBn), supporting the view that the two jets are essentially symmetrical, with an underlying jet in the range ~150-170 m/s.  Secondly, the manifestation of this jet at the surface is normally suppressed by the presence of large slow-moving formations, which are probably Rossby waves. Thirdly, the loss of the large dark formations and the narrowing of the belt may have been promoted by the decline of convective rift activity. Fourthly, these changes seem to represent a reversion to the situation that existed before 1912, when such appearances were common and were followed cyclically by vigorous 'NEB Revivals'.  Indeed, just such a Revival would ensue in 2012.

_______________________________________________________

## 1.  Introduction

Paper I [ref.1] described the cyclic pattern of NEB expansion events (NEEs)* that have occurred every 3-5 years in recent decades and also in some earlier decades.  Even earlier, in 1893-1915, similar events occurred probably every 3 years but were even more striking [refs.2&3]; the NEB narrowed considerably (the NEBn edge typically receding to only 11-12ºN), and then revived, in a process that was sometimes observed as a vigorous outbreak with many striking spots.  These events have been termed 'NEB Revivals' [refs.2&3]; they showed many similarities with the better-known SEB Revivals [refs.3&4], although the latter appear to be more highly organised.  One final NEB Revival occurred in 1926; thereafter only the more modest NEEs occurred.

> ***Footnote:***
> ***Abbreviations of belts etc.:***
> Abbreviations for belts and associated phenomena are as follows (those for the belts being standard):  NEB, North Equatorial Belt; NEDF, NEBs dark formation; NEE, NEB expansion event; NTB, North Temperate Belt; SEB, South Equatorial Belt; SED, South Equatorial Disturbance.  Suffix n or s indicates the north or south edge of the belt, and by extension the jet associated with it.

In 2011, the NEBn began receding again and we expected the next in the 3-5-yearly series of NEEs.  However, the recession then continued until the NEBn edge reached 12.5ºN in early 2012, leaving a narrow southerly NEB which also became somewhat fainter than usual.  The NEB had never been so narrow or so faint since the 1920s.  This appeared to set the scene for a NEB Revival, which indeed followed in 2012, and will be described in Paper III.

As this was the first such phenomenon in modern times, we were able to record fine details and associated phenomena that were unobservable a century ago.  The recession of the belt was preceded by disappearance of the usual NEBs dark formations (NEDFs), and of NEB rifts, and acceleration of the NEBs jet to unprecedented super-fast speeds.  I argue that there was a close association between all these phenomena, which are analogous to those accompanying SEB Fade/Revival cycles.

This paper begins by considering the structure of the NEBs jet, as inferred from ground-based and spacecraft observations, and its emerging similarity with the other two similarly fast prograde jets on the planet.  Section 2 describes the changes in its observed speed from 2008 to 2012, and Section 3 discusses the implications for the jet structure and variability, and for the role of NEDFs.  In Section 4, I describe the NEB Fade as it developed in 2011/12, and discuss how the rifts, NEDFs, and Fade/Revival cycle may be related, in the NEB as in the SEB.  Paper III will describe the subsequent NEB Revival and the concurrent NTB Revival in 2012.

The data are from our BAA/JUPOS analysis from 2008 to 2012, as in our reports on-line [ref.5], especially our final report for 2011/12 [ref.6], including its Appendix 2 on the super-fast speeds. (Figures 2-6 herein are adapted from that 2011/12 report.) A preliminary account has also been given in [ref.7].

Conventions and abbreviations are as in Paper I.  Speeds are quoted both as DL1 (degrees per 30 days relative to System I, as measured), and as $u$ (m/s relative to System III, referenced to latitude 7.0ºN).  Thus, $u = -$ (DL1—221) x 0.47763.  South is up in all figures.

*1.2.  The three fast prograde jets:  NEBs, SEBn, and NTBs*

The NEBs jet at 7ºN, called the North Equatorial Current in the older visual literature, is Jupiter's best-known jet stream.  It almost always carries conspicuous visible NEBs dark formations (NEDFs: see Paper I), also called hot spots. However, it is still unclear what is the normal wind speed of the NEBs jet peak at the visible main cloud-top level (~0.3-0.7 bars)  The speed of the NEDFs (~105 m/s) has usually been the speed observed for all features in this latitude, large or small; and yet, they are almost certainly waves embedded in a faster jet. With the improvement in amateur techniques since 2000, we have indeed detected faster speeds in most years since then, around ~120 m/s (Paper I), for smaller projections which may represent the highest-frequency class of NEDFs.  We argued that these fast speeds may be close to the true cloud-top wind speed.  However, some analyses suggested that the true wind speed is in a

'super-fast' range (~140-150 m/s), based on the speed-vs-spacing correlation for NEDFs [ref.8] and observations from Cassini [refs.9-11], as we will discuss in Section 3.  The Galileo Probe entered a large NEDF in this latitude and discovered even faster wind speed (~170-175 m/s) below the 3-bar level [refs.12&13].

Moreover, the corresponding jet at 7ºS (SEBn) always shows fast speeds at cloud-top level, averaging ~155 m/s when there are no large-scale formations on it, as is often the case. At other times it carries just one large formation, the South Equatorial Disturbance (SED), and when this is most active it suppresses the peak speeds to its east and to a lesser extent at all longitudes, giving a speed range of ~120-147 m/s on the SEBn jet [refs.14&15]. The SED, like the NEDFs, may be a Rossby wave [refs.15&16].

So why are these two jets not more symmetrical?  Can very fast speeds ever be detected at the surface of the NEBs?  Or do the large dark formations on the NEBs suppress the surface expression of the deeper jet speed?  The 2011-12 results will shed light on these questions.

The third very fast jet is on the North Temperate Belt south edge (NTBs, 23.7ºN), which has two very different states.  In the normal state, it carries vortices at DL1 ~ -60 deg/mth ($u$ ~ 125 m/s for this latitude), and the peak wind speed of the jet at cloud-top level is only ~10 m/s faster [refs.17&18].  In the 'super-fast' state, which held from 1970 to 1990 and again from 2007 onwards, it undergoes spectacular outbreaks initiated by one or more brilliant white plumes (DL1 ~ -160 deg/mth, $u$ = 170 m/s for this latitude), comparable to $u$ ~ 180 m/s recorded between such outbreaks by Voyager.  Theoretical modelling of these NTBs jet phenomena [refs.19&20] requires the jet below cloud-top level to have a permanent super-fast speed.  The cloud-top speed is suppressed by vortices when they are present.  This scenario is supported by our observations in recent years, as we will explain in Paper III.  In 2008-2012 the jet was undergoing changes in apparent cloud-top speed which would come to a head in a great NTBs jet outbreak at the same time as the NEB Revival in spring, 2012 (Paper III).

## 2.  Observations of super-fast features on NEBs, 2008-2012

Here we describe our results from measurements of amateur images in visible light from 2008-2012, as reported in BAA/JUPOS reports posted on our web site [ref.5].

```
Table 1:  All our records of super-fast speeds (-DL1 > 50):

Year          Main NEDFs       Rapid spots & projections:
              No.    DL1       Average/Consensus (Range)
                     (deg/mth) DL1 (deg/mth)        u3  (m/s)

2008          3      +4        ( -35 (-28 to -40)   122 (119 to 125)
                               ( -60 (-45 to -66)   134 (127 to 137)

2010          5      +13       ( -40 (-30 to -49)   125 (120 to 129)
              -->              ( -68 (-57 to -78)   138 (133 to 143)
              6-8    +13 to +39  --

2011          0      --        -71 (-36 to -95)     139 (123 to 151)

The right-hand half of this Table lists both fast speeds (u ~ 120 m/s)(Paper
I) and super-fast speeds (u ~ 140 m/s). They are not completely separated;
speeds in the range u ~ 125-130 m/s can be assigned to either class.
```

In 2008 July, while large formations were replaced by smaller fast projections around half the planet, the remaining large NEDF in another sector was replaced by even faster features. It became very tenuous in May and disappeared in June. In July, many small, tenuous, very oblique projections/festoons were recorded here, with white spots in EZ(N) between them, moving with average DL1 = -60 (range ~-45 to -66 deg/mth) (spacing ~12º longitude). This was faster than ever before observed on the NEBs, amounting to 134 m/s. [ref.5a]

In spring 2009, the NEBs was still devoid of large NEDFs and all the drifts recorded were in the fast, but not super-fast, range. This state ended dramatically in July when a vigorous rift system in the NEB, expanding around the planet and associated with the ongoing expansion of the NEB to the north, apparently induced the formation of many dark spots on NEBs with positive DL1, including large ones which developed into a regular array. However, they did not last through the next year. [Ref.5b]

In summer 2010, only five major NEDFs remained – long low plateaux – and these too were subsiding. As the major formations diminished, a few dark features appeared with DL1 = -29 to -36 deg/mth; and then, some much faster. The first super-fast one, in July, was a small projection/festoon with DL1 = -78 [ref.5c]. In August, the last slow-moving projection in one long sector disappeared, leaving only small tenuous festoons. At least some of these were moving with DL1 = -75, and the trend of other points on the chart suggests that this was the dominant drift rate throughout the whole sector, up to Sep. Overall, from July to Dec., we obtained 9 drifts ranging from -57 to -78: mean DL1 = -68 (±8) ($u$ = 138 m/s), mean lat. +7.0 (±0.2) ºN. A zonal wind profile obtained from amateur images on 2010 Sep.4 confirmed this very high speed, $u$ = 145 (±5) m/s [ref.21; Figure 1]. However, from mid-Sep., more normal and slow drifts (DL1 ~ +8 to +39) took over most longitudes again, as low blue-grey humps reappeared on NEBs, followed in Oct. by more substantial (though still unstable) NEBs formations. Thus, the super-fast drifts only applied where there were no normal NEBs formations.

In the 2011 apparition, the disappearance of the normal NEDFs resumed and the NEBs became completely taken over by super-fast speeds (**Figs 2 & 3**). From 2011 June to Oct., there were only 2 or 3 slow-moving features at any one time, which were small blue-grey NEBs projections with DL1 = +26, accelerating to DL1 ~ +15 to +11 in their final stages. All other tracks were super-fast: 113 tracks gave a mean DL1 = -70.6 (±14.4) ($u$ = 139 m/s). These features had the typical appearance of small dark blue-grey NEBs projections with festoons (**Figs 2 & 4)**; only their speed was exceptional.

Detailed study of the JUPOS chart (**Fig.3**) revealed even more remarkable aspects of these speeds, summarised in **Fig.5**. The speeds were gradually accelerating during 2011, and they were modulated by the few remaining slow projections, although these were very small. New super-fast projections generally appeared on the p. (east) side of these slow projections, usually with DL1 ~ -49 (±5) deg/mth, then they accelerated or were replaced by faster-moving projections, with DL1 ~ -76 (±10) deg/mth. These accelerations were usually abrupt, and occurred ~70º p. the slow projections (range, 60-95º). Sometimes the track split at this point. In several cases, though, the super-fast projections oscillated in motion during the transition, accelerating to DL1 ~ -90 before settling down to -78; three of them performed a complete cycle of oscillation with P ~ 20-30 days.

After all the slow projections had disappeared, the super-fast projections continued to accelerate, and in 2012 Jan-Feb., had a mean DL1 = -83 deg/mth (u = -145 m/s), and maximum of -95 deg/mth (u = 151 m/s).

The latitudes for all these dark spots in 2011/12 were the same as for normal dark formations, thus: slow spots, 7.7ºN (±0.2, SD); super-fast spots, 7.5ºN (<0.1, SEM); oscillating super-fast spots, 7.5 ºN (±0.2, SD). For the oscillating spots, there may have been a weak oscillation with latitude between 7.1—7.7ºN, in the sense that latitude was highest when relative longitude was lowest; i.e. latitude would correlate with acceleration, not speed. However we cannot be confident of the significance of this suspected variation.

Nor was there any correlation of the speed and spacing of super-fast features. Where regularly spaced groups could be identified, all groups with speeds of DL1 from -54 to -82 ($u$ = 131 to 145 m/s) had spacings of 11-13º, while one group moving at DL1 = -45 (127 m/s) had a spacing of 8º. As we found for the chevrons on SEBn [ref.15], there was no general correlation.

## 3. Discussion: The super-fast speeds on NEBs

The super-fast speeds (~140 m/s) were much faster than anything reported in this latitude before 2000, except for a single record from the Hubble Space Telescope (HST). Most observations from HST, like those from Voyager, showed only the speeds typical of the NEDFs [ref.22]. The only detection of super-fast speeds in HST images before 2008 was a jet peak of 150 (±10) m/s in images taken on 1994 July 29 [ref.23].

Since we reported the super-fast motions in 2008 [ref.5a], two groups [refs.9&10] have analysed HST images from that year, and confirmed that super-fast speeds were widespread: the jet peak speed was variously determined by them as 131, 143, and 140-155 m/s, over large sectors.

Now we have shown that in 2011-12, the super-fast motions (~140 m/s) extended all around the NEBs, with the disappearance of the normal NEDFs and all other slow-moving features.

### 3.1. *What are the super-fast features?*

The super-fast features looked like typical small-to-medium-sized projections [Fig.2]. The projection/festoon/chevron shape is not diagnostic of their physical nature: this shape is to be expected for various kinds of disturbance at the peak of the jet. Nevertheless, they are probably not at a much deeper level than the usual NEDFs, given their similar appearance, and the fact that they exist when there is substantial white cloud cover over the EZ. (As with the NTB in 2003 [ref.18], detection of faster speeds coincided not with loss of bright cloud cover but with loss of large dark features.)

Are the super-fast projections waves of the same type as the normal NEDFs? This seems unlikely, firstly because they do not fit on the speed-vs-spacing relationship that holds for NEDFs (Paper I), and secondly because the arguments we have made for comparable features on the SEBn jet ('chevrons') [ref.15] can also apply here. We argued that Rossby waves correspond to the NEDFs (as explained in Paper I), and the SED, and the oscillating pattern of chevrons in the Cassini movies, but that the SEBn chevrons themselves are unlikely to be Rossby waves, since they would have such low phase speed relative to the peak jet wind flow that the waves would be absorbed. Instead, the SEBn chevrons were interpreted as inertia-gravity waves, and the super-fast NEBs features could be the same kind of wave. These are waves with vertical oscillation, similar to waves on the surface of water or in terrestrial clouds downstream of mountains but with a lateral component due to the rotation of the planet; their speed increases with wavelength. On a planetary scale, a wide spectrum of these waves can be generated by a vigorous disturbance and can radiate out in all directions. For the SEBn (or NEBs), the most obvious possible sources would be the large-scale convective 'rifts' in the SEB (or NEB); but when these are absent, as in 2010 or 2011 respectively, long-wavelength waves could still impinge on the equatorial jets as omnipresent 'noise' from disturbances in even higher latitudes and, perhaps more importantly, from deep down in the underlying convection zone (M. McIntyre, personal communication). The inertia-gravity waves in some frequency ranges could be sheared and focussed into the fast eastward jet, achieving higher amplitude as they were trapped there, so possibly becoming visible as a chain of chevrons (M. McIntyre, pers.com.). Their speed would most likely be close to the true cloud-top wind speed, although they could be somewhat faster. Thus, the inertia-gravity wave model may provide a natural explanation of the chevron pattern on these jets, and it suggests that the motion observed is indeed close to the true wind speed.

### *3.2. Has the cloud-top wind speed increased?*

One striking aspect of this change is that the NEBs has taken on the same appearance, dynamics, and speed, as the equivalent jet at 7ºS (SEBn). This is further evidence that the two jets are essentially symmetrical, with potential peak speed of ~145-155 m/s at the cloud-tops. This suggests that it was the presence of large NEDFs which prevented the super-fast jet speed appearing at cloud-top level, similar to the behaviour of the SEBn jet (suppressed by the SED – refs.14&15) and NTBs (suppressed by vortices – refs.18-20). A similar argument has been made in [ref.9]: "...we propose that Rossby wave activity [the NEDFs] is responsible for the zonal [speed] variability. Removing this variability, we find that Jupiter's equatorial jet is actually symmetric relative to equator with two peaks of ~140–150 m/s located at latitudes 6ºN and 6ºS and at a similar pressure level."

But does this mean that the deep super-fast jet only spreads up to the surface when there are no NEDFs to disrupt it? –or, that the super-fast speed is always present at the surface but usually not visible? The latter possibility was raised by the Cassini imagery in 2000 which showed super-fast motions for small and subtle features despite the presence of large NEDFs. At that time there were 7 large NEDFs, with some long gaps between them, and the EZ was largely covered in unusually thick white clouds [ref.24]. In Cassini's images at 756 nm which penetrate most deeply into the clouds, possibly down to ~3 bars, Li *et al.* [ref.25] detected speeds of u = 142-175 m/s for small bright clouds within the NEDFs. These Cassini near-IR images have been reanalysed [refs.9-11] and all these studies found widespread super-fast speeds: large sectors at ~140-150 m/s, and some up to ~170 m/s, largely between the NEDFs and sometimes overlapping them. The more distinct ones were dubbed white 'scooter clouds' [ref.11]. The animated maps posted by Choi *et al.* [ref.11] appear to show a near-continuous jet with speeds in this range, not limited to certain sectors, except that the 'scooter clouds' tend to disappear (evaporate?) within the NEDFs. They note examples where clouds that have spilled off the NEB seem to mask deeper clouds with the characteristic motions of the northern EZ.

Therefore, two alternative models can describe the true wind speed of the jet at cloud-top level (~0.3—0.7 bar ), and we cannot yet decide between them.

**Model 1:** The true wind speed is ~140-150 m/s (super-fast) at all times, but distinct features do not form at that speed until NEDFs disappear (as advocated [refs.9&11] from the Cassini data). The main evidence for this is as follows.

i) The Cassini observations, described above.  However, it is possible that these small and subtle features with super-fast motions were different from, and deeper than, the ones that we have detected in more recent years.  They were generally smaller and of lower contrast than in HST 2008 data [ref.9], and were best tracked on the near-IR images. Nevertheless, a minority could be tracked in visible-light images, and were at about the usual cloud-top level.
ii) The analysis of NEDFs as Rossby waves [ref.8], which found 140 m/s to be the best fit for the speed of the jet.  But this modelling may not be conclusive, given the unknown parameters, and especially given that we have extended the linear relationship up to higher wave-numbers (Paper I), whereas Rossby wave theory predicts a dependence on $1/n^2$.

**Model 2:**  The true wind speed at cloud-tops in normal times is ~120 m/s (fast features), but permanent super-fast winds at greater depth spread upwards when the NEDFs disappear.  The main evidence for this is as follows.
i)  The Galileo Probe detected a speed of 95 (±20) m/s when it started recording near the 0.5 bar level [refs.12&13].  However, the Probe descended into a large NEDF, where there is thought to have been a profound down-welling from the adjacent jet, stretching the atmosphere up to six-fold down from cloud-top level  [ref.s.26&27]. Choi *et al.* [ref.11] pointed out that the wind speeds may also have been stretched downwards, so the speed of ~150 m/s detected at the 3-bar level [refs.12&13] may actually be the normal cloud-top speed outside the NEDF.
ii)  The presence of fast speeds and not super-fast speeds, when NEDFs are present.   But this could be because NEDFs influence the type of smaller features that can develop, rather than the cloud-top wind speed.
iii)  Our observations of continuous increase in speed during 2011 and a gradient of speed p. the surviving small NEDFs (see next section).  This is not easy to reconcile with the Model 1.

### *3.3.  Analogy with SEBn/SED:*

There are especially intriguing similarities between the NEBs features and SEBn chevrons in their behaviour just east of slower-moving features: residual NEDFs, or the SED [refs.14&15], respectively.  This was first suspected in 2008 July, when the NEBs strikingly resembled the SEBn [ref.5a]. There were only three reasonably large and persistent NEDFs, forming a group with DL1 ~ +4 deg/mth, analogous to the SED; preceding them were smaller 'fast' projections with DL1 = -28 to -40 deg/mth, analogous to the disturbed sector p. the SED; and further p., more tenuous small projections and white spots with mean DL1 ~ -60 deg/mth.  Thus the pattern of speeds formed a gradient remarkably similar to that on the SEBn.  Given the limited extent of the phenomenon, it was impossible to tell whether it was anything more than chance.  Indeed, when super-fast speeds reappeared in 2010,  no such gradient was seen. But a gradient became unmistakable when the widespread super-fast speeds developed in 2011.

In 2011, new projections generally appeared on the p. (east) side of the two remaining slow projections, usually with DL1 ~ -49  deg/mth, then they accelerated abruptly or were replaced by faster-moving projections, with DL1 ~ -76 deg/mth, starting ~70º p. the slow projections.  This behaviour was just like that of chevrons on SEBn p. the SED [refs.14&15].

It is remarkable that such a powerful effect was exerted by such small slow-moving features; they were so visually inconspicuous that they might not have been recognised at all but for their tracks on the JUPOS chart (Fig.3).  The potency of this effect can explain why super-fast speeds were not observed on the NEBs previously.  Even a very small NEDF reduces the speeds over tens of degrees; no doubt, the usual large NEDFs suppress them totally.

It was also intriguing to observe that several of the super-fast projections underwent oscillations during their acceleration to full speed.  This was also reminiscent of the behaviour of chevrons on SEBn in the Cassini movies [ref.15]. It is not obvious whether the phenomena are comparable, because the oscillations on the SEBn were principally in latitude, whereas on the NEBs they were principally in longitudinal speed. Although coupled latitudinal oscillations were suspected, the phase relationship between latitude and speed is unexpected and may be difficult to explain. Another difference is in the period of the oscillations: 6.7 d on the SEBn, but ~20-30 d on the NEBs in 2011. It is worth noting that large NEDFs occasionally oscillate with similar period: in 1999/2000, we reported that several oscillated with P ~ 20 d.

## 4. Shrinkage and fading of the NEB in 2011/12 [ref.6]

In 2011, the whole NEB was very quiet, and its north edge was gradually receding after the last expansion event in 2009. Exceptionally, no rifts were present inside it at all after 2011 July (when there was just one small southerly rift). There were however 6 remarkably dark barges (cyclonic circulations), 3 of which were also very large and conspicuous.

The recession of the NEBn edge is shown in Fig.6. Up until August, the NEB spanned 8.5º to 17ºN. In Sep., sectors of the northern half began to fade (brighten), until in Nov. all of the northern half was lighter. So far, this appeared to be normal for this stage of the NEB cycle; but what happened next was remarkable. The northern half continued to fade rapidly until by late Jan. it was just a very pale fawn colour, almost white, leaving just a narrow southerly NEB between 9 and 12.5ºN, which had also become somewhat fainter than usual. The narrowing was also evident in methane images [ref.6 Figures 2-4].

Six dark barges had all persisted since 2010, and were extremely dark brown from 2011 Aug. to Dec., while the NEBn had cleared around them. These were visually the most prominent features on the planet, and observers commented that they had never seen barges so dark. Thereafter they were not quite so dark, but they all persisted until March without further fading.

There were also five white ovals in the NTropZ, although hard to see because of the whiteness around them. Three of these had been prominent white ovals in the previous apparition, including long-lived white spot Z, and the other two had been reddish anticyclonic dark spots.

The JUPOS chart tracked many dark spots retrograding at 16-18ºN, mostly with DL2 ~+9 deg/mth, but with DL2 = +17 to +43 deg/mth in one sector. These features were tiny projections on NEBn, moving with the retrograding NEBn jet. Unusually, the maximum speed (+43 deg/mth) was similar to the maximum speed observed by spacecraft, perhaps made visible now by the absence of confounding disturbance around it.

## 5. Discussion: Relationship between NEB rifts, NEDFs, and NEB fading

*A reversion to pre-1912 behaviour of the NEB?*
The NEB had not been so narrow nor faint since the 1920s [ref.3]. From 1893 to 1912, it regularly shrank to this narrow state or even narrower, prior to a vigorous outbreak of spots leading to revival every 3 years, which were grander versions of the familiar NEEs, usually being vigorous affairs much like an SEB Revival [ref.3 chapter 8.5]. We therefore suspected in 2011-12 that the exceptional narrowing of the NEB was setting the scene for a NEB Revival. Indeed, a NEB Revival began suddenly in 2012 March and proceeded vigorously during solar conjunction (Paper III). So in 2011-12 we have observed the first major 'NEB Fade' in modern times.

It can hardly be a coincidence that this 'NEB Fade' coincided with three other exceptional phenomena: unprecedented super-fast speeds, and the disappearance of both rifts and NEDFs.  I suspect that in its quiescent state, the NEBs generally has super-fast speeds of ~140-150 m/s just like the SEBn.  But such speeds were not detected a century ago because visual observations did not track the small tenuous festoons that have now revealed them.  The linkage between the NEB Fade and the disappearance of rifts and NEDFs is comparable to the fading of the SEB, which likewise begins with the complete disappearance of the usual rifting in the SEB [refs.28&29]. A connection is also plausible given the converse behaviour in a NEE, when a new northerly rift often appears to initiate the NEB expansion, and expanding rifts affect the behaviour of NEDFs (Paper I).  Nevertheless, rifts and NEDFs do not disappear before an ordinary NEE (Paper I); we can only consider full NEB Fades.

For NEB Fades, of course, the observational record is limited. Visual observations a century ago could readily detect major rifts, but were insufficient to distinguish active rifts from quieter light patches, or to determine systematically whether rifts were absent.  So the linkage between rift disappearance and NEB narrowing/fading can only be established for 2011-12.  NEDFs are much more readily observable than rifts, and when NEB Fades and Revivals were happening frequently before 1912, the NEBs edge generally had few or no NEDFs, except during the Revivals [ref.3 chapter 8.5].  But conversely in 1925 (preceding the last NEB Revival, in 1926), plenty of large NEDFs remained on the NEBs even while the NEB became very narrow [ref.30].

Is some level of rift activity is required to sustain NEDFs? We know that rifts can trigger the reappearance or expansion of NEDFs [refs.1&3].  However NEDFs can decline without loss of rifts.  In 2008-09, for example, during the previous narrowing of the NEB, small mid-NEB rifts had appeared repeatedly, although rifts were absent in the hemisphere without NEDFs.  But rifts were still present throughout 2010; and in 2011, the NEDFs were already waning before the last NEB rifts disappeared in 2011 July.

The suggested connections could be valid even if imperfect. NEDFs, as Rossby waves, may well be self-sustaining for some time without external stimulation.  And rift activity is only assessed visually [refs.1&2] and may not be a quantitative marker for relevant levels of energy release in the NEB.

It is evident that these connections do not provide a complete scenario for the process leading to the NEB Fade, particularly as there was a tendency for NEDFs to disappear from 2008 onwards, intermittently reversed by resurgence of rift activity. Nevertheless, I conjecture that the unique NEB Fade of 2011-12 was initiated in summer 2011 by a decline of convective activity in the NEB which resulted in the complete loss of both NEB rifts and NEDFs, and thus to the acceleration of the NEBs jet, and eventually to the enhanced narrowing of the NEB.

## Acknowledgements

This work depends entirely on the images produced by numerous observers around the world, especially those in 2011-12 who are listed alongside our report [ref.5].  I am also very grateful to Marco Vedovato, for producing regular maps from the images, and to all of the JUPOS team (G. Adamoli, M. Jacquesson, M. Vedovato, H-J. Mettig) who measured the images to derive the drift rates.

*Figures (miniature copies)*

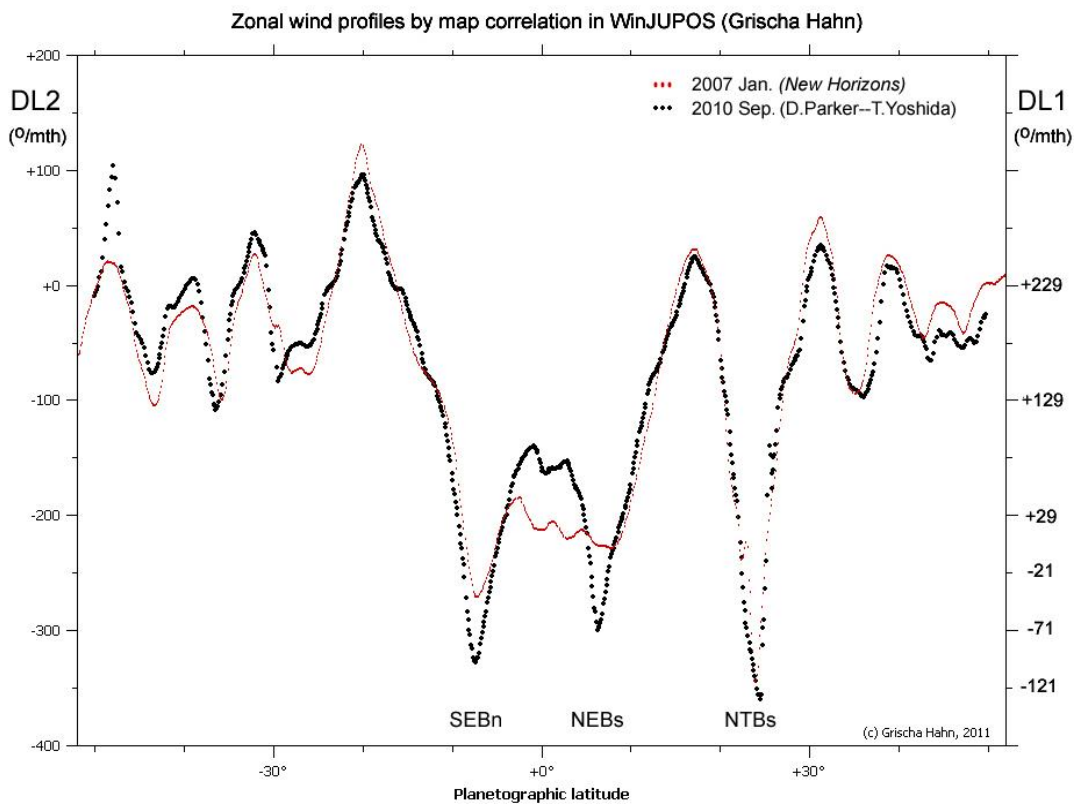

**Fig.1.** Zonal wind profiles in 2007 Jan. (New Horizons) and 2010 Sep. (from amateur images by D. Parker and T. Yoshida), made by Grischa Hahn using his map correlation function in WinJUPOS (from ref.21). Of the three fast prograde jets, those on SEBn and NEBs were much faster in 2010 than 2007, while the NTBs had super-fast speed in both years (shortly before the outbreaks of 2007 and 2012 respectively).

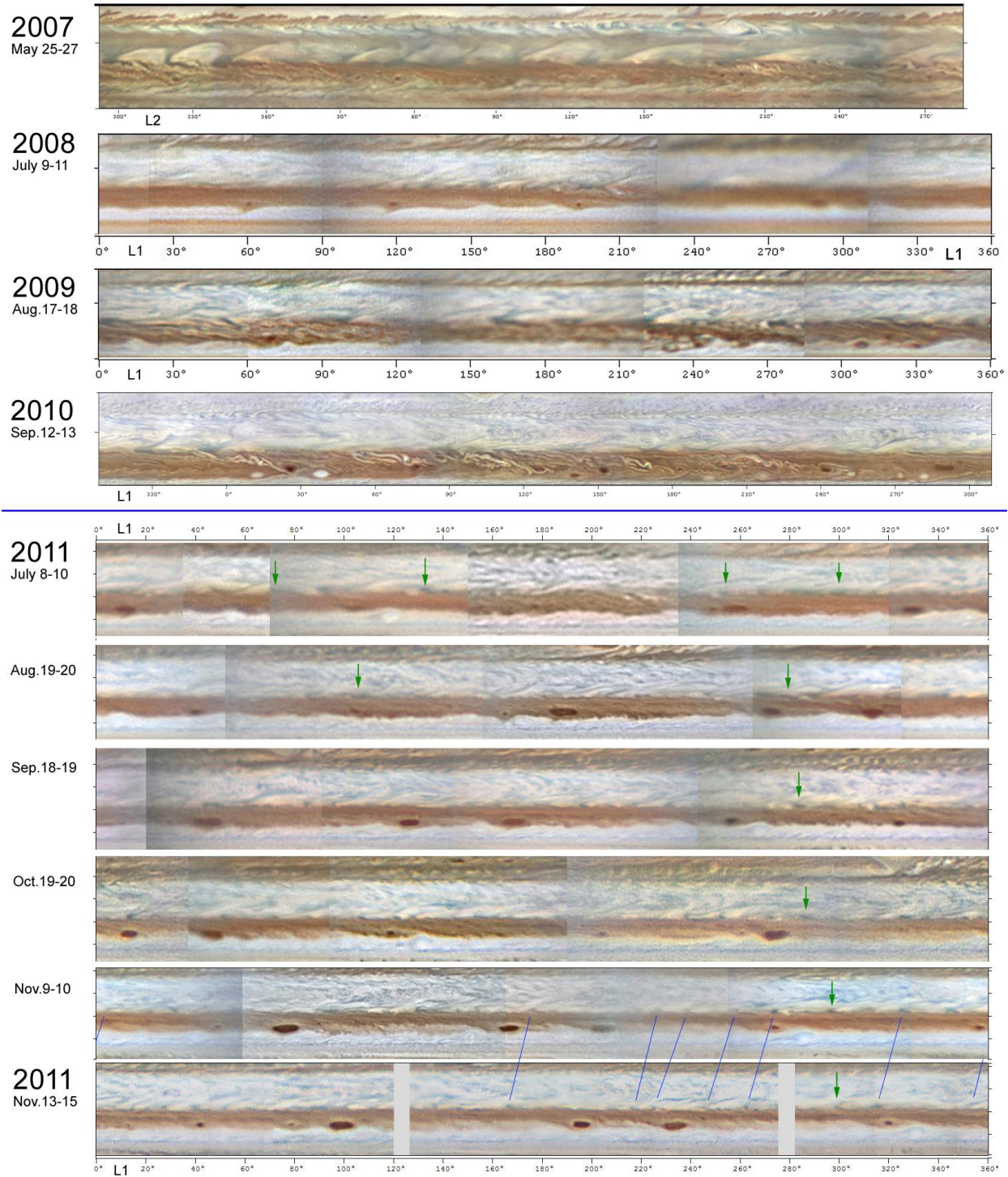

**Fig.2.** Maps of the NEB around the whole planet, from 2007 to 2011 (reduced from ref.6 Fig.22). *South is up in all figures.*

The 2011 maps are aligned in L1. The few remaining normal-speed projections, identified from the JUPOS charts, are indicated by green arrows; there is nothing to visibly distinguish them from the super-fast ones. Individual super-fast projections are connected by blue lines on the last two maps, only a few days apart.

For 2007 and 2010, maps were made by Damian Peach from his own images. (The 2007 map was made in L2 so there may be discontinuities in the equatorial region.) For other years, maps were made by Marco Vedovato from images by the following observers: (2007-2010) C. Go, X. Dupont, T. Barry, T. Akutsu; (2011) D. Peach, T. Akutsu, T. Kozawa, T. Yoshida, B. Combs, S. Saltamonti,, E.Morales Rivera, I. Sharp. J. Beltran Jovani, D. Kolovos, W. Jaeschke. The final map was made by Freddy Willems from his own images.

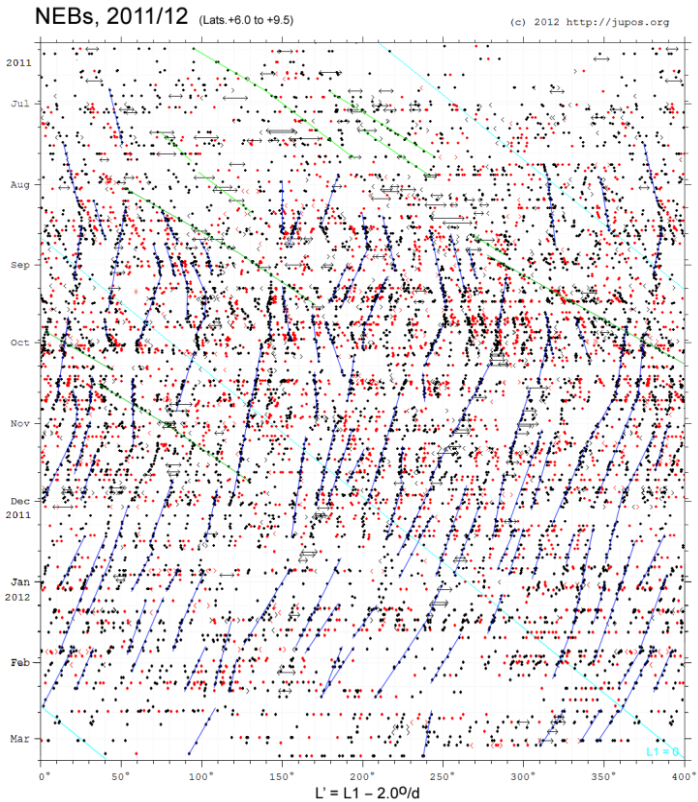
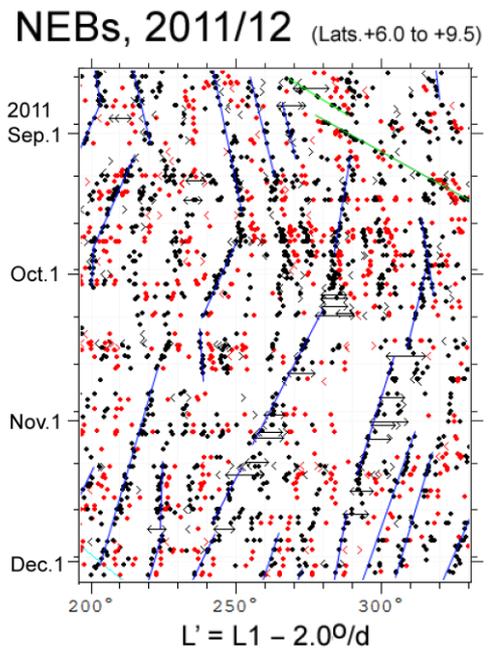

**Fig.3.** JUPOS chart of the NEBs in 2011/12, in a longitude system moving at -2.0 deg/day relatuve to System I. (A) The whole chart in miniature; (B) excerpt at full scale. (The whole chart at full scale is available with ref.6). Black points are dark features (NEBs projections); red points are white spots. Green lines indicate normal-speed projections; blue lines indicate super-fast projections.

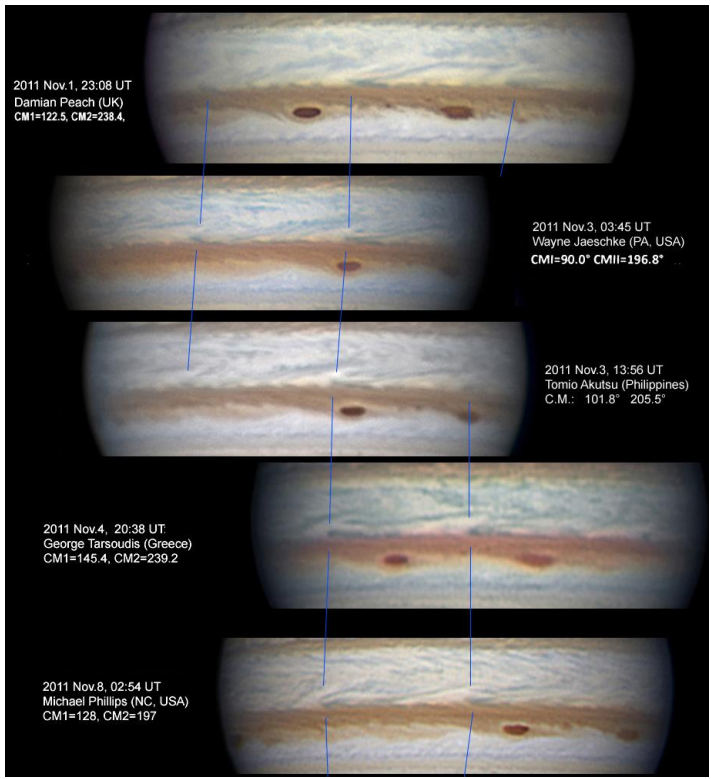

**Fig.4.** Hi-res images tracking some super-fast NEBs projections in 2011. (Excerpt from ref.6 Fig.23)

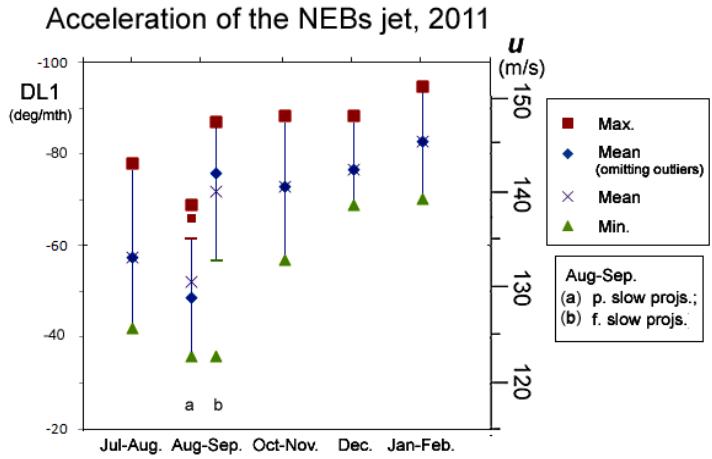

**Fig.5.** Chart of speed ranges vs time for the NEBs projections in 2011/12. For Aug-Sep., the projections are divided into those preceding the few remaining slow projections (a) and those following them (b), and a few outliers (individual values very different from most of the group) are shown separately. (Ref.6 Fig.24)

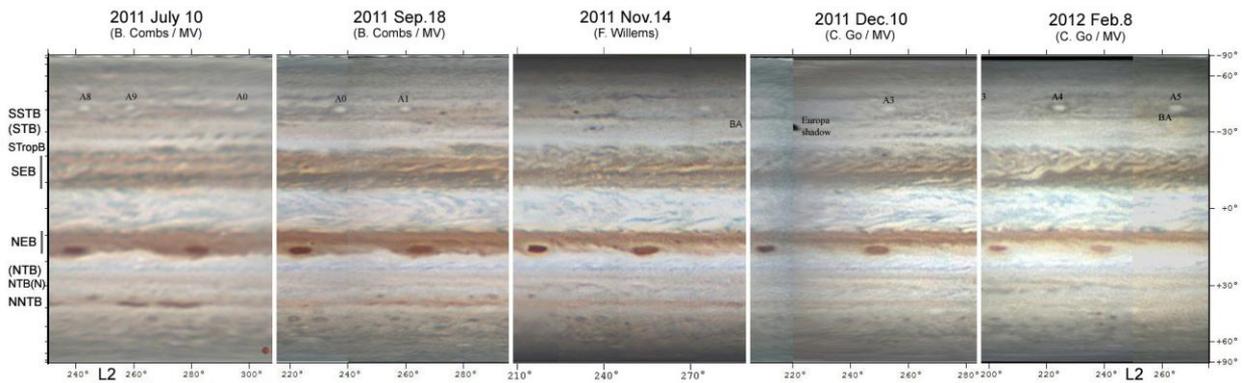

**Fig.6.** Portions of maps in 2011-12, showing the progressive narrowing and fading of the NEB. (Ref.6 Fig.26)